\documentclass[aps,prb,twocolumn,superscriptaddress,amsmath,amssymb]{revtex4}
\usepackage{graphicx}
\usepackage{epsfig}
\usepackage{float}
\usepackage{multirow}
\usepackage{dcolumn}
\usepackage{bm}

\begin{document}

\hsize=6.5 true in
\vsize=8.9 true in
\parindent =15 pt
\hoffset= 0.1 in

\title{ Theoretical Predictions of Superconductivity in Alkali Metals under High Pressure }

\author{  Lei Shi  }
\affiliation{ School of Computational Sciences, George Mason University, Fairfax, VA 22030}
\author{  D. A. Papaconstantopoulos }
\affiliation{ School of Computational Sciences, George Mason University, Fairfax, VA 22030}
\affiliation{ Center for Computational Materials Science, Naval Research Lab, Washington, DC 20375}

\begin{abstract}
We calculated the superconductivity properties of alkali metals
under high pressure using the results of band theory and the rigid-muffin-tin theory
of Gaspari and Gyorffy. Our results suggest that at high pressures Lithium,
Potassium, Rubidium and Cesium would be superconductors with
transition temperatures approaching $5-20 K$. Our calculations also suggest
that Sodium would not be a superconductor under high pressure even if
compressed to less than half of its equilibrium volume. We found that the 
compression of the lattice strengthens the electron-phonon coupling through
a delicately balanced increase of both the electronic and phononic components
of this coupling. This increase of the electron-phonon coupling in Li is
due to an enhancement of the $s$-$p$ channel of the interaction, while in the 
heavier elements the $p$-$d$ channel is the dominant component.  
\end{abstract}

\maketitle

\section{Introduction}

Neaton and Ashcroft~\cite{neaton} predicted
that at high pressures Li forms a paired ground state.
Subsequently, Christensen and Novikov~\cite{christensen} showed that $fcc$ 
Li under increased pressure may reach a superconducting transition 
temperature \( T_{c} \)=$50-70 K$. This prediction has been supported by 
Shimizu et al~\cite{shimizu} and Struzhkin et al~\cite{struzhkin} who reported
measurements of superconductivity in compressed Li with a
\( T_{c} \) ranging from $9 K$ to $20 K$. The overestimate of $T_{c}$ reported in 
Ref[2] is probably due to an error by these authors as we discuss
in Sec. III. Using the rigid-muffin-tin approximation(RMTA) formulated
by Gaspari and Gyorffy~\cite{gaspari}, we performed
calculations for two other alkali metals, K and Rb,
and predicted that at high pressures they both would be superconductors
with transition temperatures approaching $10 K$~\cite{k_rb}. Before these
works, the only known superconducting alkali metal was Cs, which
becomes superconducting above 7.0 GPa with a transition temperature of
about $1.5 K$.~\cite{wittig} Similar theoretical results for Li, 
K and Cs were also reported by Ashcroft~\cite{ashcroft}, Profeta et al.~\cite{profeta}, 
Kasinathan et al.~\cite{kasinathan}, Tse et al~\cite{tse} and Stocks et al.~\cite{stocks} 

In this paper, we have extended our study to 
Li, Na and Cs using again the RMTA. Our calculations demonstrate
that Li and Cs display superconductivity at high pressure above 
15 GPa and 3.5 GPa respectively. Our results also showed the lack of 
superconductivity for Na up to 90 GPa. We compare these new calculations
with our previous results of K and Rb to show a complete picture of 
superconductivity properties of alkali metals under high pressure.  
Our calculations indicate that the $s$-$p$ channel of contribution to the 
Hopfield parameter $\eta$ dominates Li under high pressure, while the $p$-$d$ 
channel contribution is the major reason that K, Rb and Cs become 
superconductors under high pressure.  

The massive structural phase transitions of all alkali metals have be extensively
investigated during the past few decades.~\cite{liu,xie,tse,young,hanfland,
schwarz,katzke,deemyad} All alkali metals were found to be stable with $bcc$ structure
under ambient pressure and will transform to the $fcc$ structure at about 20, 65, 11.5, 7,
and 2.3 GPa pressure for Li, Na, K, Rb and Cs respectively. In this paper, our
investigations are focused on the $bcc$ and $fcc$ structures within the above pressure
ranges which correspond to volume changes as large as $40\%$ from equilibrium.
The conclusions we draw in this paper are valid for the
structure which is stable in the experimental range of pressure.

Tomita et al~\cite{tomita} recently reported experimental results for Li, Na
and K. They confirmed the superconductivity of Li above 20 GPa at temperatures
reaching 15 $K$. They also pointed out the absence of superconductivity in Na and K 
for pressure up to 65 and 43.5 GPa respectively without specifying whether the 
measurements were extended to the $fcc$ phase. However, for K, our calculations and those 
of Profeta et al~\cite{profeta} show superconductivity in the $fcc$ phase. Further
experimental work may be required.

\section{ Theory and computational details}

McMillan's strong coupling theory~\cite{mcmillan} defines an electron-phonon coupling constant by:

\begin{equation}
\lambda=\frac{\eta}{M<\omega^2>}=\frac{N(\epsilon_{F})<I^2>}{M<\omega^2>}
\label{eq_lambda}
\end{equation}

\noindent where \( M \) is the atomic mass; $\eta$ is the Hopfield parameter~\cite{hopfield};
$N(\epsilon_{F})$ is the total density of states(DOS) per spin at the
Fermi level, $\epsilon_{F}$; $<$$I^2$$>$ is the square of the electron-ion
matrix element at $\epsilon_{F}$; and $<$$\omega$$>$ is
the average phonon frequency. The $<$$I^2$$>$ is determined by
the RMTA approximation given by Gaspari and Gyorffy~\cite{gaspari}(GG) formula:

\begin{align}
<I^2>=\frac{\epsilon_{F}}{\pi^2}\sum_{l} 2(l+1) \frac{\sin^2(\delta_{l+1}-
\delta_{l})}{N_{l}^{(1)} N_{l+1}^{(1)}} \frac{N_{l}N_{l+1}}{N^2(\epsilon_{F})}
\label{isquare3}
\end{align}

\noindent where $\delta_{l}$ are scattering phase shifts, \( N_{l}(\epsilon_{F}) \) is the \( l \)th
component of the DOS per spin, and \( N_{l}^{(1)} \) is the free
scatterer DOS.

There are two important papers that address the question of the spherical approximation in the GG theory.
The first one by John~\cite{john} shows that the GG formula is exact for
cubic systems with one atom in the unit cell and with $l$+1 up to 2. This validates our
results for Li and Na where the $p$-$d$ and $d$-$f$ channels have negligible contributions. In
addition, Butler et al.~\cite{butler} have obtained a generalization of the
GG formula for $l$$>$2 that is exact for systems in which all atoms sit at sites having cubic
symmetry. Their expression contains certain cross terms which they showed have small  or
canceling contributions to the value of $\eta$. Therefore, our calculations for K, Rb, and
Cs which have strong diagonal contributions from $p$-$d$ but not $d$-$f$ scattering are also
reliable.

In order to determine the quantities entering Eq.~\ref{isquare3}, we performed Augmented 
Plane Wave (APW) calculations of the band structures and total energies of the targeted 
alkali metals in the local density approximation(LDA) following the Hedin-Lundqvist 
prescription~\cite{hedin} in both $bcc$ and $fcc$ structures, for a wide range of pressures. 
Since the LDA underestimates the equilibrium lattice constant of alkali metals by
approximately $3\%$ we applied a uniform shift to our results for $E(V)$ so that at
the experimental equilibrium volume we have the minimum energy. To accomplish this
the pressure shifts are 1.5, 0.4, 0.4, 0.4 and 0.25 GPa for Li, Na, K, Rb and Cs respectively.
We also used the APW results to determine the pressure variation of the bulk moduli $B$.
The $B$ values were used to compute the average phonon frequency which has been taken
to be proportional to the product of bulk modulus and volume~\cite{jarlborg}:

\begin{equation}
<\omega^2> = CB(V)V^{1/3}
\label{omega_BV}
\end{equation}

\noindent where the constant C is determined by the $<$$\omega^2$$>$ and the Debye temperature
$\Theta_{D}$ under normal pressure using the formula:

\begin{equation}
\omega^2=\frac{1}{2}\Theta^2_{D}
\label{eq_omega}
\end{equation}

The approximation made in using Eq.~\ref{omega_BV} and Eq.~\ref{eq_omega} gives 
reasonable values of $\lambda$ and leads us to believe that our results should differ only
quantitative from direct phonon evaluations via linear response theories.
The transition temperature for superconductivity is given by the McMillan
equation~\cite{mcmillan}:

\begin{equation}
T_{c}=\frac{<\omega>}{1.2} exp[\frac{-1.04(1+\lambda)}{\lambda-
\mu^*(1+0.62\lambda)}]
\label{mcmillan}
\end{equation}

\noindent where $\mu^*$ is the Coulomb pseudopotential, that we have set equal
to 0.13.



\section{Results}

\begin{figure}[h]
\centering
\includegraphics[width=3.0in,height=2.9in]{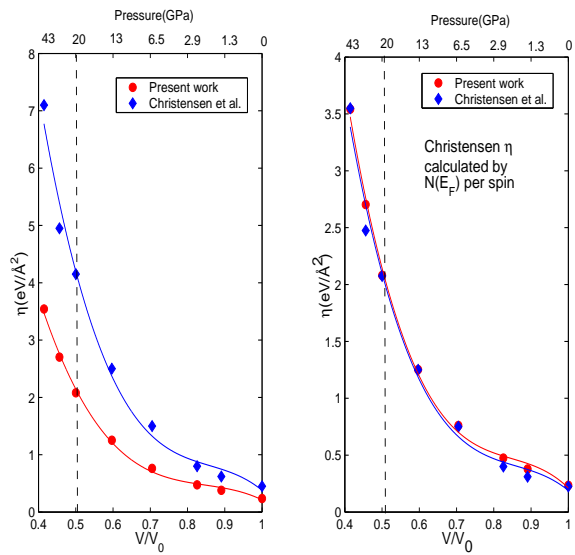}
\caption{ Comparison of $\eta$ between our results and Christensen and Novikov for $fcc$ Li in the
left panel, and the corrected result in the right panel. The vertical dashed line indicates 
the transition from $bcc$ to $fcc$.   }\label{eta_our_chris}
\end{figure}

Before we present our detailed results, we wish to bring to the attention of the reader
that in our opinion there must be an error in the evaluation of $\eta$ for Li by
Christensen and Novikov.~\cite{christensen} The error is that in Eq.~\ref{eq_lambda} 
they must have used $N(\epsilon_{F})$ 
for two spins instead of one as is the correct implementation of the Mcmillan
theory. Fig.~\ref{eta_our_chris}(left panel) demonstrates a clear discrepancy between our
results and those of Ref 2. Dividing the $\eta$ of Ref 2
by two gives a perfect agreement with our results as shown in Fig.~\ref{eta_our_chris}(right panel)
and as we discuss below lowers the value of $\lambda$ and brings $T_{c}$ close to the measured
value.

\begin{figure}[ht]
\begin{minipage}[c]{.26\textwidth}
\centering
\includegraphics[width=1.7in,height=2.8in]{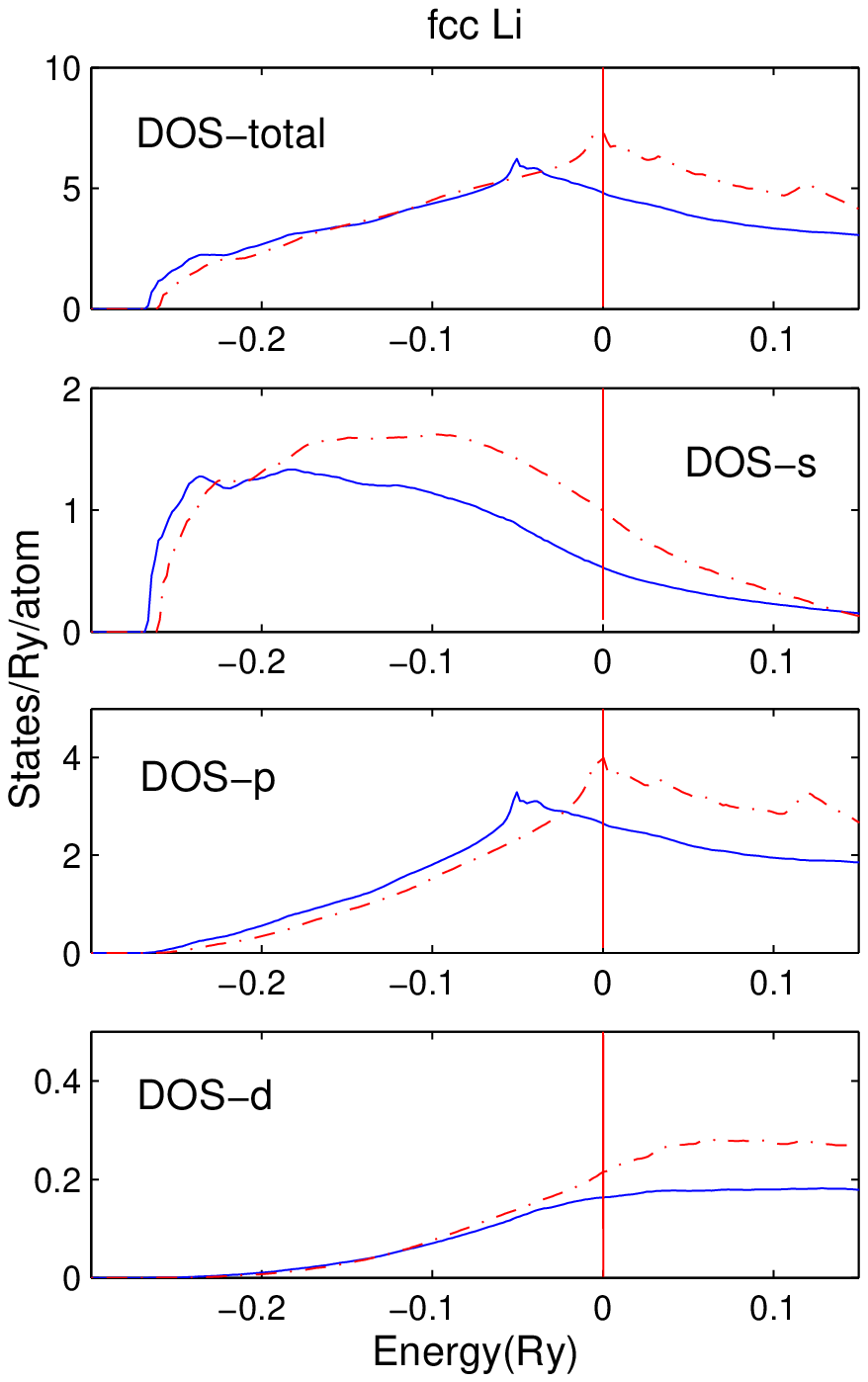}
\end{minipage}%
\begin{minipage}[c]{.26\textwidth}
\centering
\includegraphics[width=1.7in,height=2.8in]{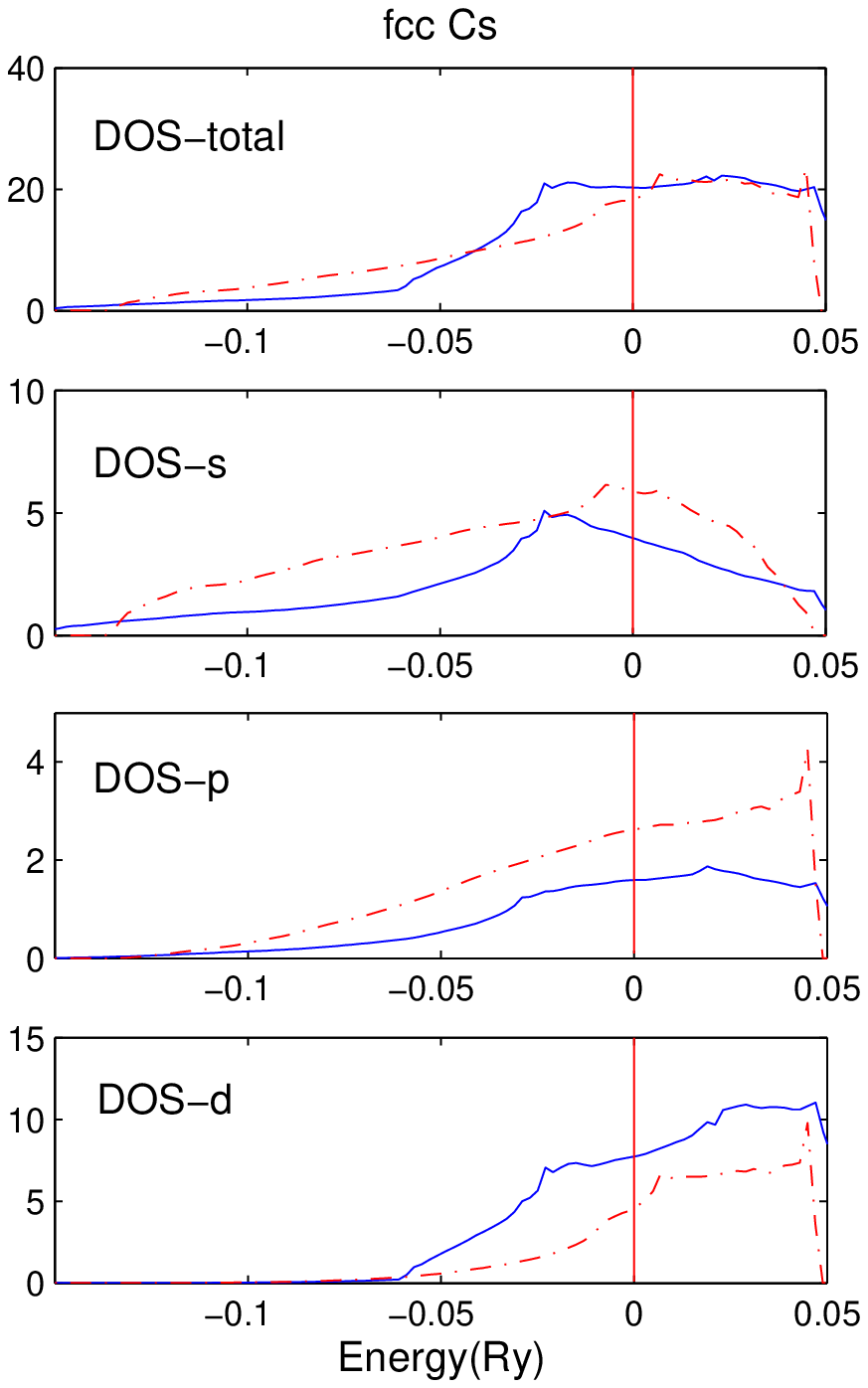}
\end{minipage}%
\caption{ Total density of states and angular momentum decompositions of $fcc$ Li and Cs under 
normal and high pressure. The solid and dashed lines denote the high and ambient 
pressure respectively. }
\label{cs_dos}
\end{figure}

To demonstrate
the effect of high pressures on the density of states at the Fermi level for alkali metals 
we plotted the total density of states and its angular momentum decompositions for $fcc$
Li and Cs under ambient and high pressures in Fig.~\ref{cs_dos}. For Li, we note that both $N(\epsilon_{F})$
and its angular momentum decompositions decrease under pressure. On the other hand, for Cs,
we observe a decrease of the $s$ and $p$-like density of states with a remarkable increase of the 
$d$-like density of states $N_{d}(\epsilon_{F})$ under high pressure. This increase of the 
\( d \) density of states at $\epsilon_{F}$ makes the largest contribution to the large value of 
$\eta$ at large pressure as discussed below. We also observed the same trend as in Cs for 
the alkali metals K and Rb. In contrast, for both the $bcc$ and $fcc$ phases of 
Li and Na, the density of states decreases as the lattice is compressed. 
This fact gives a partial explanation of the absence of superconductivity under high
pressure for Na. For Li, however, the increase in the matrix 
elements $<$$I^2$$>$ is strong enough to ensure the overall large increase in $\eta$, which 
we will discuss later. 

\begin{figure}[h]
\centering
\includegraphics[width=3.0in,height=2.9in]{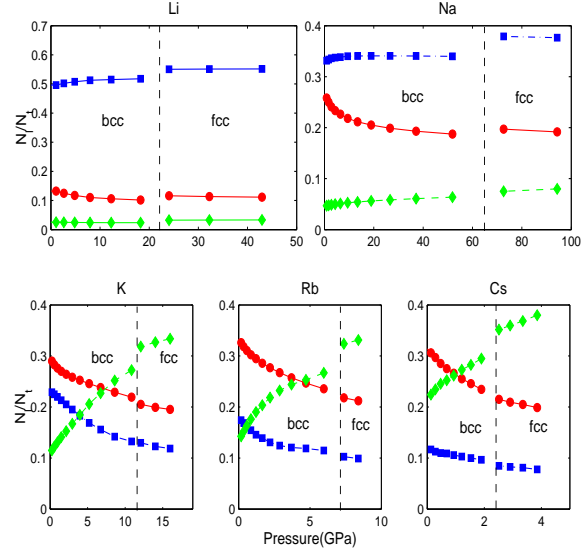}
\caption{ Angular momentum decomposed DOS divided by the total DOS at $\epsilon_{F}$,
the filled square, circle and diamond symbols denote $s$, $p$ and $d$ states respectively. }\label{spd5}
\end{figure}

\begin{figure}[h]
\centering
\includegraphics[width=3.0in,height=2.9in]{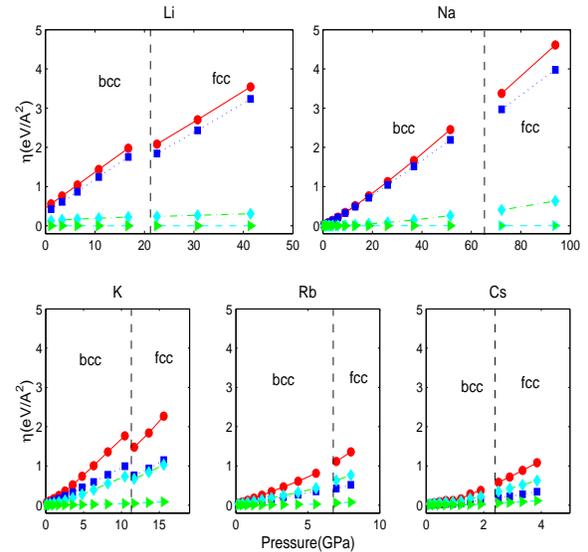}
\caption{ Total $\eta$ and its contributions from the $s$-$p$, $p$-$d$ and $d$-$f$ 
channels, the filled circle, square, diamond and triangle symbols denote total, 
$s$-$p$, $p$-$d$ and $d$-$f$ chanels respectively.}\label{spd_eta}
\end{figure}

We show the ratios 
\( N_{l}(\epsilon_{F})/N(\epsilon_{F}) \) as a function of pressure in Fig.~\ref{spd5}
where the vertical line indicates the pressure where the transition from 
$bcc$ to $fcc$ occurs in the experiments. 
These ratios are crucial in the determination of $\eta$.
It is important to note that the ratio \( N_{d}(\epsilon_{F})/N(\epsilon_{F}) \)
of the heavier alkali metals K, Rb and Cs increases rapidly as a function of
increasing pressure. The buildup of the $d$-like DOS under high pressure
causes the large values of $\eta$ for the heavier elements under high pressure shown 
in Fig.~\ref{spd_eta}. For Li and Na, the $d$ ratio is very small and the $s$
ratio dominates. We note that for Na, the $p$ ratio decrease with increasing
pressure. 

In Fig.~\ref{spd_eta}, we show the total $\eta$ and its contributions from 
$s$-$p$, $p$-$d$ and $d$-$f$ scattering channels as a function of pressure in 
both the $bcc$ and $fcc$ structures. \( \eta \) in all
five alkali metals in the $bcc$ and $fcc$ lattices increases significantly 
with increasing pressure. Among all alkali metals,
Li has the second largest increase of $\eta$ with increasing pressure, which is
one of the reasons why Li is a superconductor under high pressure.
Fig.~\ref{spd_eta} shows that different elements have different scattering channels
as the dominant contributors to the total $\eta$. More specifically for
Li and Na, the $\eta$ contribution of the $s$-$p$ channel increases rapidly
with increasing pressure while the other two contributions
stay around zero. For K and Rb, when pressure is increased the $\eta$
contributions from both the $s$-$p$ and $p$-$d$ channels increase
quickly. For Cs, the $p$-$d$ channel contributes the major increase
of $\eta$ with pressure increasing. For all elements, the $\eta$ contribution
of $d$-$f$ channel is always very close to zero. The graph also demonstrates that
the largest portion of the $\eta$ increase in Li and Na under high pressure is contributed 
by the $s$-$p$ channel. Na shows the largest value of $\eta$ among all the alkali 
metals when pressure reaches 90 GPa. However, as we discuss below Na has a very large value
of $<$$\omega$$>$ which lowers significantly the value of $\lambda$.

\begin{figure}[h]
\centering
\includegraphics[width=3.0in,height=2.9in]{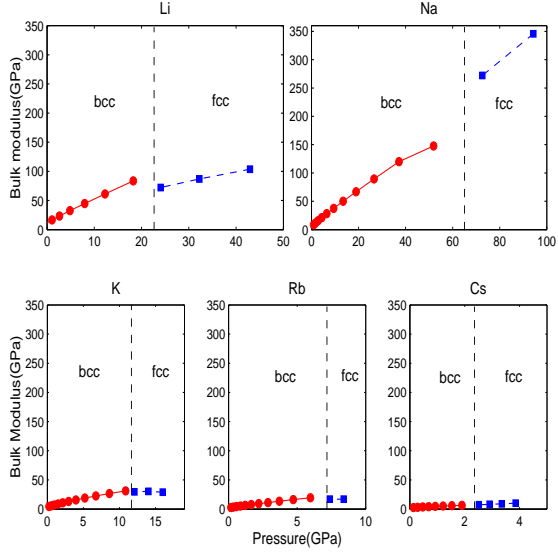}
\caption{ The bulk moduli $B$ as a function of pressure, 
the filled circle and square symbols denote $bcc$ and $fcc$ respectively. }\label{bulk}
\end{figure}

\begin{figure}[h]
\centering
\includegraphics[width=3.0in,height=2.9in]{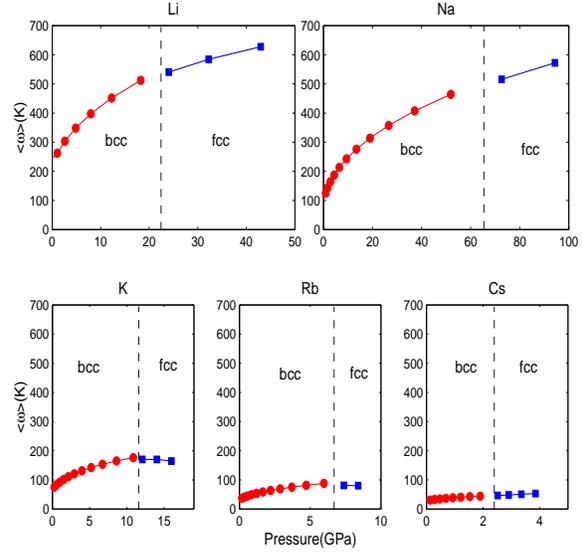}
\caption{ Average phonon frequency $<\omega>$ as a function of pressure. 
The filled circle and square symbols denote $bcc$ and $fcc$ respectively.}\label{omega}
\end{figure}

\begin{figure}[h]
\centering
\includegraphics[width=3.0in,height=2.9in]{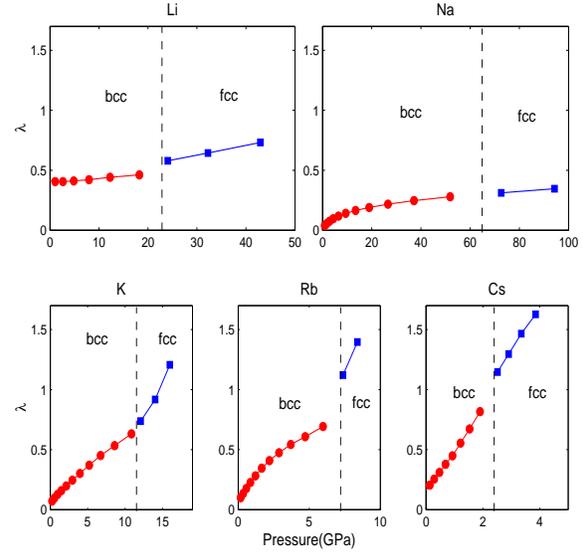}
\caption{ Electron-phonon coupling constant
$\lambda$ as a function of pressure. The filled circle and square
symbols denote $bcc$ and $fcc$ respectively.}\label{omegalambda}
\end{figure}

\begin{figure}[h]
\centering
\includegraphics[width=3.0in,height=2.9in]{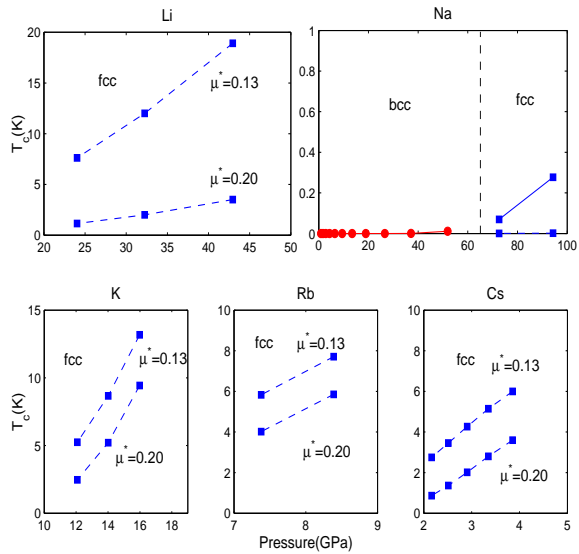}
\caption{ Transition temperature $T_{c}$ as a function of pressure,
the filled circle and square symbols denote $bcc$ and $fcc$ respectively. }\label{tc}
\end{figure}

Fig.~\ref{bulk} shows
the increase of the bulk modulus $B$ as a function of pressure. We
note that the bulk moduli of Li and Na increase dramatically with increasing
pressure. The bulk modulus increase of K, Rb and Cs is around $1/3$ to $1/10$ of
that for Li and Na. Fig.~\ref{omega} shows the prefactor,
$<$$\omega$$>$, in the McMillan equation. We note that the $<$$\omega$$>$ 
for Li and Na, in both $bcc$ and $fcc$ structures, increases much faster than in the other 
alkali metals. This is the other reason that Li has larger $T_{c}$ than
K, Rb, and Cs despite the fact that $\lambda$ of K, Rb, and Cs can be larger 
than that of Li under pressure. Combining the values of $\eta$ and $M$$<$$\omega^2$$>$,
we determined the electron-phonon coupling constant $\lambda$ using
Eq.~\ref{eq_lambda}. In Fig.~\ref{omegalambda}, we show 
$\lambda$ as a function of pressure. It is evident that for the alkali metals
Li, K, Rb, and Cs, under pressure, $\lambda$ reaches large values suggesting
that these metals could display superconductivity. To quantify
our predictions for superconductivity, we have calculated \( T_{c} \) for all 
five elements in both the $bcc$
and $fcc$ structures at high pressure using the McMillan equation with
Coulomb pseudopotential values \( \mu^*=0.13 \) and 0.20. Fig.~\ref{tc} shows the 
$T_{c}$ of the
five alkali metals in the $fcc$ structure as a function of pressure.
As one might expect from the large values of the electron-phonon interaction $\lambda$, the
elements Li, K, Rb, and Cs are predicted to be superconductors with $T_{c}$ larger than 5 $K$
for pressures higher than 20, 11, 7, and 3.5 GPa respectively.
Clearly our results
for the value of $T_{c}$ are sensitive to the value of $\mu^{*}$ and increase of $\mu^*$ to 0.2
suppresses $T_{c}$ by a factor of four for Li and about $20\%$ to $40\%$ for K, Rb
and Cs, as shown in Fig.~\ref{tc}.

Our calculations suggest that Na does not
display superconductivity because the electron-phonon coupling constant
$\lambda$ remains small (\( \lambda \approx 0.25 \)) for compressions up
to 90 GPa as shown in Fig.~\ref{omegalambda}. The absence of superconductivity for
Na is determined by delicate balance of the increasing electronic and phononic
components of the electron-phonon coupling constant. 
Na has a $\eta$ which increases faster than in K, Rb and Cs,
but slower than in Li. However, the increase of $\eta$ in Na is mostly canceled out
by its $M$$<$$\omega^2$$>$ which increases much more rapidly than in K, Rb and Cs. 
Therefore, Na does not display
strong enough electron-phonon coupling to become a superconductor under high
pressure up to 90 GPa in both the $bcc$ and $fcc$ structures. 

\section{CONCLUSIONS}
As it can be seen from Eq.~\ref{eq_lambda},
the increase of $\lambda$ may be caused by either an increase of $\eta$ or a decrease 
of $M$$<$$\omega^2$$>$, or a combination of these two factors. For K, Rb, and Cs, 
because the increase of the bulk 
modulus under pressure is slow, the increase of $\eta$, which is mainly caused by an 
increase of $N(\epsilon_{F})$, dominates the determination of $\lambda$. Therefore, 
K, Rb, and Cs have a large $\lambda$ under high pressure, which results in the
prediction of superconductivity. The increase of $\lambda$ in Li under pressure 
is due to the increase of $<$$I^2$$>$ rather than $N(\epsilon_{F})$, which is actually decreasing
under pressure, and overcomes the increase of $M$$<$$\omega^2$$>$ which in fact helps 
by increasing the prefactor $<$$\omega$$>$ in the $T_{c}$ equation. Therefore, Li would 
be a superconductor under high pressure. Na is different from Li and from the other alkali 
metals because its large $\eta$ is canceled by the increasing bulk modulus under pressure 
which reduces the value of $\lambda$. Therefore, a delicate balance of $\eta$ and 
$M$$<$$\omega^2$$>$ results in Li being a superconductor but not Na. Our calculated $T_{c}$
for Li showing in Fig~\ref{tc} is in very good agreement with experiment.~\cite{shimizu}  

It should be mentioned here that Tse et al~\cite{tse} argued that the softening
of the $TA$ phonon branch near the $X$ point causes the occurrence of superconductivity
in Li. Our view is that this is an oversimplification
and is not a sufficiently quantitative explanation. On the other hand the RMTA takes an average
in all directions on the Fermi surface and not just in one high symmetry point. 
Regarding the mechanism of superconductivity in K, Rb, and Cs we suggest that this
is due to the increased $d$-like character of the wave functions at \( \epsilon_{F} \) 
at high pressures. This enhanced $d$ character near $\epsilon_{F}$ has also been
documented by other authors~\cite{louie} and by experiment~\cite{badding}. We used it 
here to justify the RMTA which is successful in
transition metals as shown by Papaconstantopoulos et al.~\cite{papa_super}

{\bf Acknowledgment:} We wish to thank Dr. Michael J. Mehl for valuable discussions
and comments.

\end{document}